\documentclass[aps,floatfix,onecolumn,a4paper,noshowpacs, nofootinbib,superscriptaddress,11pt]{revtex4}
\usepackage{graphicx,float}\usepackage{graphicx,float}
\usepackage[all]{xy}
\usepackage{amsmath,upgreek}
\usepackage{amssymb}
\usepackage{color}
\usepackage{epsfig,bm}		
\usepackage{graphicx,epstopdf}
\usepackage{subfigure}
\usepackage{pdfpages,slashed}
\usepackage[colorlinks,hyperindex]{hyperref}

\definecolor{green1}{RGB}{0,128,0}
\hypersetup{backref=true,pagebackref=true}
\hypersetup{%
  colorlinks = true,
  linkcolor  = blue,
  citecolor = cyan,
}
\usepackage{bookmark,textgreek}
\usepackage{hyperref,color,xcolor}
\hypersetup{hidelinks,hyperindex=true,colorlinks=true,breaklinks=true,urlcolor= blue}
\hypersetup{%
  colorlinks = true,
  linkcolor  = blue
}\usepackage{amssymb}
\newsavebox{\foobox}

\usepackage{graphicx,float,tikz}
\usepackage[all]{xy}
\newcommand\ringring[1]{%
  {% make an Ord atom
   \mathop{\kern0pt #1}\limits^{% set a box over the variable
     \vbox to-1.85ex{
       \kern-2ex % lower the ring accents
       \hbox to 0pt{\hss\normalfont\kern.1em \r{}\kern-.45em \r{}\hss}%
       \vss % fill
     }% end of \vbox
   }% end of the superscript
  }% end of \mathop
}\newcommand\orcidgayane{{\href{https://orcid.org/0000-0002-0785-2826}{\orcidicon}}}
\newcommand{\orcidicon}{%
	\begin{tikzpicture}
	\draw[lime, fill=lime] (0,0)
		circle [radius=0.16]
		node[white] {{\fontfamily{qag}\selectfont \tiny ID}};
	\draw[white, fill=white] (-0.0625,0.095)
		circle [radius=0.007];
	\end{tikzpicture}	\hspace{-2mm}
}
\newcommand{\bpartial}{\mathop{\partial\kern -4pt\raisebox{.8pt}{$|$}}}

\newcommand{\bes}{\begin{subequations}}
\newcommand{\ees}{\end{subequations}}
\def\beq{\begin{eqnarray}}

\def\eeq{\end{eqnarray}}
\def\be{\begin{equation}}
\def\ee{\end{equation}}

\begin{document}

\title{Nuclear configurational complexity in high-energy hadron-hadron scatterings}
\author{G. Karapetyan\!\orcidgayane}
\email{gayane.karapetyan@ufabc.edu.br}
\affiliation{Federal University of ABC, Center of Mathematics, Santo Andr\'e, 09580-210, Brazil}

\begin{abstract}
This paper investigates high-energy hadron-hadron scattering utilizing  AdS/QCD correspondence, gravitational form factors, and the Brower--Polchinski--Strassler--Tan pomeron exchange kernel.
We use the configurational complexity to estimate the slope of the total cross section for hadron-hadron interactions at the high-energy regime.
Our approach agrees well with the phenomenological cross sections regulated by Pomeron exchange involving the pion-nucleon, nucleon-nucleon, and pion-pion, in data from the TOTEM collaboration (LHC).
In the case of pion-nucleon and pion-pion scattering, the agreement for the global critical point of the configurational complexity has good accuracy within $2.3\%$.
\end{abstract}
\maketitle
\section{Introduction}
The Color Glass Condensate (CGC) approach is a powerful and practical tool for examining the mechanism of high-energy processes in QCD.  Configurational entropy (CE) techniques in this setup forecast the key points that aid in the comprehension of the vast amounts of experimental data collected by various cooperative groups, mainly involving hadron spectroscopy in AdS/QCD and the quark-gluon plasma (QGP)   \cite{daRocha:2024sjn,daRocha:2024lev,Casadio:2023pmh,Karapetyan:2023sfo,daRocha:2023nsb,daRocha:2022bnk,Braga:2022yfe,Barreto:2022mbx,Casadio:2022pla,Karapetyan:2022rpl,Barreto:2022len,Barreto:2022ohl,daRocha:2021jzn,daRocha:2021ntm,Karapetyan:2018yhm,daRocha:2021imz, Fernandes-Silva:2018abr,Ferreira:2020iry, Bernardini:2019stn,Bazeia:2021stz,Braga:2016wzx,Braga:2023qee,Braga:2021fey,Braga:2021zyi,Braga:2020hhs}. 
The nuclear CE approach to nuclear systems in extreme conditions can be used to derive relevant information about different aspects of hadron-hadron or photon-hadron cross sections \cite{Karapetyan:epjp,Karapetyan:plb,Karapetyan:2020epl, Karapetyan:2021epjp,Karapetyan:2023kzs,Karapetyan:2021crv}. This method has been effectively applied to characterize nuclear reactions \cite{Ferreira:2019inu,Bernardini:2018uuy,Braga:2018fyc,Bernardini:2016hvx,Barbosa-Cendejas:2018mng,Ferreira-Martins:2021cga,Toniato:2025gts}, glueballs \cite{Bernardini:2016qit}, quarkonia \cite{Braga:2017fsb}, the QGP \cite{daSilva:2017jay}, baryons \cite{Guo:2024nrf,Colangelo:2018mrt}, heavy-quark exotica \cite{Karapetyan:2021ufz,MartinContreras:2023oqs,MartinContreras:2022lxl,Ma:2018wtw}. Techniques employing the holographic entanglement entropy were also developed in Refs. \cite{daRocha:2021xwq,daRocha:2020gee,DaRocha:2019fjr}.
The CC is based on Shannon's information entropy \cite{Gleiser:2012tu, Gleiser:2011di, Gleiser:2013mga, Gleiser:2018kbq,Gleiser:2018jpd,Sowinski:2015cfa,Gleiser:2014ipa,Gleiser:2015rwa}.
Critical values of the CE have been employed to explain the phenomena involving the CFT dual to AdS black holes \cite{Meert:2020sqv,Abdalla:2009pg,Casadio:2016aum,Fernandes-Silva:2019fez,Braga:2020myi,Bazeia:2018uyg}.
Global critical values of the CC carry all information content about the stability of the complex nuclear systems. 
In the framework of the CGC, the nuclear CE has been deeply employed to compute hadronic cross sections \cite{Karapetyan:2019epl}. 
The nuclear CE can be inferred when the cross section plays the role of a localized function related to the probability of the reaction product for any nuclear spatial configuration 
\cite{Karapetyan:2017edu,Karapetyan:2016fai}. We will use in this work a complementary approach based on the configurational complexity (CC) \cite{Gleiser:2018kbq,Gleiser:2018jpd}. 
The idea behind CC consists of realizing that if the power spectral density underlying wave modes is not uniform, the inherent complexity is higher, whereas if the power spectrum is uniform, the complexity becomes higher \cite{Gleiser:2018jpd}.

Due to the many experimentally studied scattering processes that have been carried out at high energy regimes, such as the data obtained at the LHC, the inner composition of hadrons and the quark-gluon pairing have recently attracted a lot of attention. The factorization theorem allows one to break down the cross section at high energies into its hard and soft components within the framework of QCD. Although the hard part of the cross-section can be estimated using the perturbation method of QCD, the soft part is more difficult to calculate because of its non-cumulative origin. The parton distribution functions (PDFs) parametrization, which is represented by the Bjorken scaling variable $x$ and the energy scale $Q^2$, can therefore be tried using a phenomenological approach. High-energy scattering reactions have been examined in such a setting using the AdS/CFT correspondence and the holographic QCD. The Pomeron exchange at the small $x$ range is  proposed to describe the parton dynamics. Within \cite{Brower:2006ea} the deep analysis based on the gauge/string duality was conducted by Brower, Polchinski, Strassler, and Tan (BPST). This approach assumes the Pomeron exchange's potential cause of the cross sections \cite{Karapetyan:2021crv}. 

Utilizing the Pomeron exchange kernel, together with the gravitational form factors, the nucleon-nucleon, the pion-nucleon, and the pion-pion cross sections are analyzed from the point of view of the CC paradigm. The Pomeron exchange model for high-energy hadronic interactions in  holographic QCD is addressed in Sec. II. The CC-based analysis  on experimentally data of hadron-hadron cross sections are discussed in Sec. III. We conclude in Sec. IV.

\section{Holographic QCD and hadron-hadron cross sections}

To compute the hadron scattering total cross section ($\chi$), let us depict  the PST Pomeron exchange kernel. The scattering amplitude in the eikonal formulation  reads 
\begin{eqnarray}
{\cal A} (s, t) = 2 i s \int d^2 b\, \exp\left(i \bm{k_\perp} \cdot \bm{b}\right) \int dz'\,dz \textsf{P}_{13}(z) \textsf{P}_{24}(z') \left[ 1-e^{i \chi (s, \bm{b}, z', z)} \right].
\label{eq:amplitude}
\end{eqnarray}
Eq. \eqref{eq:amplitude} denotes the Mandelstam variables by $s$ and $t$, and the 2D impact parameter is taken as $\bm{b}$, whereas $z$ and $z'$ are the energy scale of QCD in the dual AdS for the incident and target particles, respectively, 
being $\textsf{P}_{13}(z)$ and $\textsf{P}_{24}(z')$ the density distributions of the two hadronic states in the AdS bulk, normalized by
\begin{equation}
\int dz \textsf{P}_{13}(z) = 1 =\int dz' \textsf{P}_{24}(z').
\label{eq:normalization_condition}
\end{equation}
The optical theorem in the single-Pomeron exchange model leads to the total cross section \cite{Watanabe}
\begin{equation}
\upsigma(s) = 2 \int d^2b \int dzdz' \textsf{P}_{13} (z) \textsf{P}_{24} (z') \Im \chi (s,\bm{b},z,z'). \label{eq:tcs_original}
\end{equation}
Hence the total hadron-hadron cross section reads \cite{Watanabe}
\begin{align}
&\upsigma(s) = \frac{g_0^2 \uprho^{3/2}}{8 \sqrt{\pi}} \int dzdz' \textsf{P}_{13} (z) \textsf{P}_{24}(z') (zz') \Im [\chi_{c}(s,z,z')], \label{eq:tcs_with_CK}
\end{align} for 
\begin{align}
&\Im [\chi_c(s,z,z') ] \equiv\frac{1}{\sqrt{\tau}} \exp\left[{(1-\uprho)\tau-\frac1{\uprho \tau}\ln ^2\left(\frac{z}{z'}\right)}\right], \label{eq:CK}
\end{align}
denoting the pure imaginary part of the scattering amplitude and $\tau = \ln (\uprho z z' s/2)$ with $g_0^2$ and $\uprho$ are the total cross section's slope and overall factor, respectively. The nucleon mass ($m_N \sim 1$ GeV) is the pivotal value characterizing the nucleon-nucleon scattering process, corroborating to strong-coupled low-energy dynamics  playing a central role in QCD. Then the modified BPST kernel can be used in the following form, which has the same functional form as the conformal kernel
\begin{align}
&\Im [\chi_{mod} (s, z, z')]=
\Im [\chi_c (s, z, z') ]+ \mathcal{F} (s, z, z') \Im [\chi_c (s, z, z_0 z_0' / z') ],
\label{eq:MK}
\end{align}
where
\begin{align}
\mathcal{F} (s, z, z') &= 1 - 2 \sqrt{\uprho \pi \tau} e^{\eta^2} \mbox{erfc}( \eta ), \\
\eta &=  \frac{1}{\sqrt{\uprho \tau}}\left[-\ln \left(\frac{z z'}{z_0 z_0'} \right) + \uprho \tau\right],
\end{align}
with the coordinates cut-off parameters, $z_0$ and $z'_0$, in QCD scale, fixed by hadron masses.

The interacting hadrons' density distributions, $\textsf{P}_{13}(z)$ and $\textsf{P}_{24}(z')$, in Eq. \eqref{eq:amplitude} are described by  gravitational form factors, provided by bottom-up AdS/QCD models that use the hadron-Pomeron-hadron 3-point functions. The density distribution of the nucleon is expressed using the left-handed and right-handed components of the Dirac field ($\Uppsi_L$ and $\Uppsi_R$) via the Bessel function, since the nucleon can be expressed as a solution to the 5D Dirac equation \cite{Watanabe,Karapetyan:2021crv}
\begin{align}
&P_N (z) = \frac{1}{2z^{3}} \left[ \Uppsi_L^2 (z) + \Uppsi_R^2 (z) \right], \\
&\Uppsi_L (z) = \sqrt{2}\frac{ z^2 \mathcal{J}_2 (m_N z)}{z_0^N \mathcal{J}_2 (m_N z_0^N)}, \
\qquad\qquad \Uppsi_R (z) = \sqrt{2}\frac{ z^2 \mathcal{J}_1 (m_N z)}{z_0^N \mathcal{J}_2 (m_N z_0^N)},
\end{align}
The fixed cut-off parameter ($z_0^N=4.081$ keV) obeys the conditions of the normalization $\mathcal{J}_1 (m_N z_0^N) = 0$  for the proton and neutron  masses.
The pion wave function $\Uppsi$ is extracted from the bottom-up AdS/QCD model motion equation of mesons \cite{Watanabe}. The density distribution of the pion is written as the following:
\begin{align}
&P_\uppi (z) = \frac{ \left[ \Uppsi'(z) \right] ^2 }{4 \pi^2 f_\uppi ^2 z}  + \frac{\sigma^2 z^6 \Uppsi (z)^2 }{ f_\uppi ^2 z^3 },
\end{align}
where the prime denotes the derivative with respective to $z$ and the pion wave function reads 
\begin{align}
&\Uppsi \left( z \right) = z\Upgamma \left(\frac{2}{3} \right) \left( {\frac{\upalpha }{2}} \right)^{1/3} \Biggl[ I_{ - 1/3} \left( {\upalpha z^3 } \right) - \mathcal{I}_{1/3} \left( {\upalpha z^3 } \right)\frac{{\mathcal{I}_{2/3} \left( {\upalpha (z_0^\uppi)^3 } \right)}}{{\mathcal{I}_{ - 2/3} \left( {\upalpha (z_0^\uppi)^3 } \right)}} \Biggr]
\end{align}
 with $z_0^\uppi = 3.1051 {\rm keV}$ and $\upalpha = \frac{2 \pi}3 \sigma$, for $\sigma \approxeq (332$ MeV$)^3$, and $f_\uppi$ is the pion decay constant.
The cutoff parameter, $z_0^\uppi$, carries the $\uprho$ meson mass, $m_\uprho$, by the zeros of $\mathcal{J}_0 (m_\uprho z_0^\uppi) = 0$.

\section{Configurational complexity of the BPST Pomeron exchange kernel}

No additional adjustable parameter is required to determine the density distribution. The only two parameters that need to be taken out of the experiments in the BPST Pomeron exchange kernel approach are $g_0^2$ and $\uprho$ (see Eq. \eqref{eq:CK}. We use the TOTEM collaboration experimental data at LHC \cite{Watanabe} and the recent hadron-hadron collision data from the Particle Data Group in the energy range $10^2 < \sqrt{s} < 10^5$ GeV for our analysis. The data from Ref. has been utilized to determine the checkpoint values for the two adjustable parameters \cite{Watanabe,Karapetyan:2021crv}, given $\uprho = 0.824$ and $g_0^2 = 6.27 \times 10^2$.

\begin{figure}[h]
	\centering
	\includegraphics[width=7.8cm]{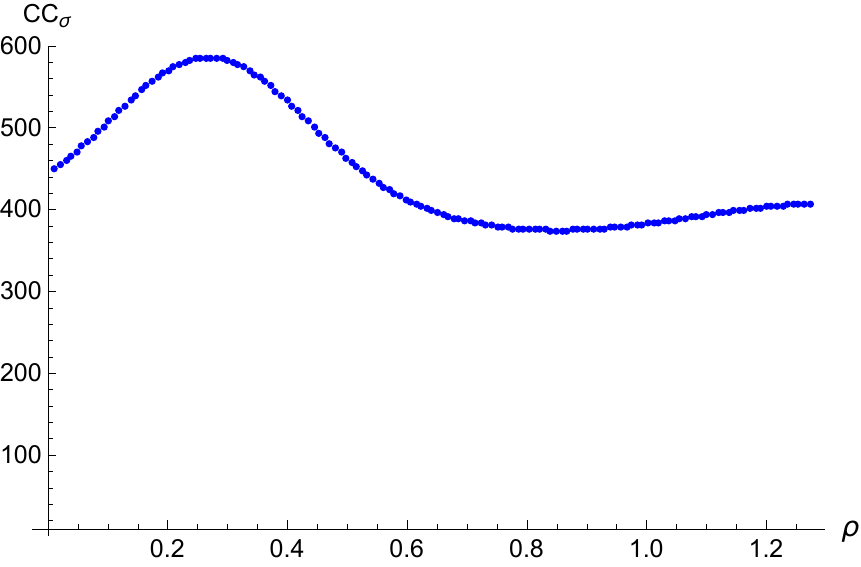}
	\caption{CC as a function of the parameter $\uprho$. Global minimum corresponds to $\uprho = 0.843$, with CC = 375.02 nat (Natural unit of information.)}
	\label{fi1}
\end{figure}

Using the Pomeron exchange kernel we compute the total hadron-hadron scattering cross-sections at a high energy range. One can presume that the Pomeron exchange approach is a universal tool to describe the process at high energies since the model's adjustable parameters are independent of the hadrons' inherent characteristics. The selected normalizable modes of the pion and nucleon both support this fact. In this manner, if the nucleon-nucleon interaction parameters are already set and exclude other movable parameters, the cross-section for the incident pion can be readily calculated. Furthermore, the constant masses of the participated hadrons characterize the scale of the scattering processes at high energies. Thus, in this scenario, the conditions for all three of the processes under investigation are also the same. which results in the total cross-section ratios between them having a constant value. For the collective coordinates, which were derived by decomposing the Fourier function into the number of weighted components for the total hadron-hadron cross sections, we employ the normalized structure factor in this paper \cite{Bernardini:2016hvx}. Since the modal fraction of the CC controls the cross section for any collision, which is the probability of the nuclear reaction, it contains all the necessary information. Conversely, the CC  can be computed, starting with the  Fourier transform of the cross section 
\begin{equation}
\label{34}
\upsigma_{}({k,\uprho})=\frac{1}{2\pi}\! \int_{\mathbb{R}}\upsigma_{} (z,\uprho)\, e^{ikz} d z.
\end{equation}
Then, the modal fraction of CC can be determined as:
\begin{equation}\label{modall}
f_{\upsigma_{}({k,\uprho})}=\frac{\vert \upsigma_{}({k,\uprho}) \vert^2}{\textsc{max}_k\vert \upsigma_{}({k,\uprho})\vert ^2}.
\end{equation}The normalization by the maximum mode yields positivity of the CC. 
The set CC reads:
\begin{equation}
\label{333}
{\rm CC}(a,b) = - \int_{\mathbb{R}} f_{\upsigma_{}({k,\uprho})} \ln f_{\upsigma_{}({k,\uprho})}d k.
\end{equation}
One obtains the CC global critical points by applying Eqs. (\ref{34}) - (\ref{333}) to the hadronic cross section. The results are plot in Fig. \ref{fi1}. 

Taking into account the total hadron-hadron scattering cross section, the CC versus $\uprho$ calculations are shown in Fig. \ref{fi1}. The nuclear CC equals 375.02 nat, as the global minimum of the nuclear configurational entropy, with the slope of the total cross section $\uprho = 0.843$ in \ref{fi1}.  For hadron-hadron interactions at high energies, our calculation's result agrees well with the value of parameter $\uprho=0.824$ found in Ref. \cite{Watanabe}, within a 2.3\%. 
At  $\uprho = 0.843$, the nuclear system 
remains at its configuration of most stability, which is usually determined  by the degrees of freedom yielding the most prevalent state upon  reaching thermodynamic equilibrium. Our results corroborate to it. 
We can still extended out results in the light of the seminal Refs.  \cite{Correa:2015lla,Bonora:2014dfa}.

\section{Conclusions}

The total cross sections for distinct hadron-hadron scattering reactions has been computed via the Pomeron exchange kernel model in AdS/QCD, as in  Refs. \cite{MarinhoRodrigues:2020yzh,MarinhoRodrigues:2020ssq}. Both the nucleon and the pion densities, given by their respective gravitational form factors, were employed. The model's premise ensures complete autonomy from the observed $s$ value due to the energy dependence of the cross section, governed by the BPST kernel. In the investigated hadron energy spectrum (100 GeV $\sqrt{s}$ < 10 TeV), the total scattering cross sections align acceptably with the data from Ref. \cite{Watanabe}. It can be inferred from the paper's findings that the BPST Pomeron exchange kernel serves as a useful instrument in the examination of high-energy hadron interactions featuring non-perturbative gluonic dynamics. The nuclear CC approach was employed, and the global minimum was assessed, indicating the critical and stable point of the hadron-hadron reactions.
We discovered a satisfactory 2.3\% agreement between the calculated parameters and those predicted by Ref. \cite{Watanabe}. 
\paragraph*{Acknowledgments:} GK thanks to The S\~ao Paulo Research Foundation -- FAPESP (grant No. 2018/19943-6).%


\begin{thebibliography}{999}

\bibitem{daRocha:2024lev}
R.~da Rocha,
%``Deformations of the AdS\textendash{}Schwarzschild black brane and the shear viscosity of the quark\textendash{}gluon plasma,''
Eur. Phys. J. Plus \textbf{139}, no.11, 1006 (2024)
%doi:10.1140/epjp/s13360-024-05795-8
[arXiv:2409.17325 [hep-th]].

\bibitem{Casadio:2023pmh}
R.~Casadio, R.~da Rocha, A.~Giusti and P.~Meert,
%``Thermodynamic and configurational entropy of quantum Schwarzschild geometries,''
Phys. Lett. B \textbf{849}, 138466 (2024)
%doi:10.1016/j.physletb.2024.138466
[arXiv:2310.07505 [gr-qc]].


\bibitem{Karapetyan:2023sfo}
G.~Karapetyan, W.~de Paula and R.~da Rocha,
%``Configurational information measure of mesonic states in 4-flavor AdS/QCD,''
Phys. Lett. B \textbf{846}, 138174 (2023)
%doi:10.1016/j.physletb.2023.138174
[arXiv:2304.05122 [hep-ph]].

\bibitem{daRocha:2023nsb}
R.~da Rocha and P.~H.~O.~Silva,
%``Configurational entropy and shape complexity of strange vector kaons in AdS/QCD,''
Eur. Phys. J. Plus \textbf{138},  729 (2023)
%doi:10.1140/epjp/s13360-023-04372-9
[arXiv:2302.05785 [hep-ph]].

\bibitem{daRocha:2022bnk}
R.~da Rocha,
%``Information entropy of nuclear electromagnetic transitions in AdS/QCD,''
Nucl. Phys. B \textbf{992}, 116232 (2023)
%doi:10.1016/j.nuclphysb.2023.116232
[arXiv:2208.07191 [hep-th]].

\bibitem{Braga:2022yfe}
N.~R.~F.~Braga, L.~F.~Faulhaber and O.~C.~Junqueira,
%``Confinement-deconfinement temperature for a rotating quark-gluon plasma,''
Phys. Rev. D \textbf{105}, no.10, 106003 (2022)
%doi:10.1103/PhysRevD.105.106003
[arXiv:2201.05581 [hep-th]].

\bibitem{Barreto:2022mbx}
W.~Barreto, A.~Herrera-Aguilar and R.~da Rocha,
%``Configurational entropy of generalized sine\textendash{}Gordon-type models,''
Annals Phys. \textbf{447}, 169142 (2022)
%doi:10.1016/j.aop.2022.169142
[arXiv:2207.06367 [hep-th]].

\bibitem{Casadio:2022pla}
R.~Casadio, R.~da Rocha, P.~Meert, L.~Tabarroni and W.~Barreto,
%``Configurational entropy of black hole quantum cores,''
Class. Quant. Grav. \textbf{40},  075014 (2023)
%doi:10.1088/1361-6382/acbe89
[arXiv:2206.10398 [gr-qc]].

\bibitem{Karapetyan:2022rpl}
G.~Karapetyan and R.~da Rocha,
%``Nuclear information entropy, gravitational form factor, and glueballs in AdS/QCD,''
Eur. Phys. J. Plus \textbf{137},  762 (2022)
%doi:10.1140/epjp/s13360-022-02952-9
[arXiv:2202.08206 [hep-ph]].

\bibitem{Barreto:2022len}
W.~Barreto and R.~da Rocha,
%``Gravitational collapse in AdS: instabilities, turbulence, and information,''
Eur. Phys. J. Plus \textbf{137}, 845 (2022)
%doi:10.1140/epjp/s13360-022-03048-0
[arXiv:2202.03378 [hep-th]].

\bibitem{Barreto:2022ohl}
W.~Barreto and R.~da Rocha,
%``Differential configurational entropy and the gravitational collapse of a kink,''
Phys. Rev. D \textbf{105}, 064049 (2022)
%doi:10.1103/PhysRevD.105.064049
[arXiv:2201.08324 [hep-th]].

\bibitem{daRocha:2024sjn}
R.~da Rocha and P.~H.~O.~Silva,
%``Deformed AdS/QCD, mesonic mass spectra, and the differential configurational entropy: Still a margin for heavier resonances,''
Phys. Rev. D \textbf{110}, no.12, 126019 (2024)
%doi:10.1103/PhysRevD.110.126019
[arXiv:2412.02375 [hep-ph]].

\bibitem{daRocha:2021jzn}
R.~da Rocha,
%``AdS graviton stars and differential configurational entropy,''
Phys. Lett. B \textbf{823}, 136729 (2021)
%doi:10.1016/j.physletb.2021.136729
[arXiv:2108.13484 [gr-qc]].

\bibitem{daRocha:2021ntm}
R.~da Rocha,
%``Information entropy in AdS/QCD: Mass spectroscopy of isovector mesons,''
Phys. Rev. D \textbf{103}, 106027 (2021)
%doi:10.1103/PhysRevD.103.106027
[arXiv:2103.03924 [hep-ph]].

\bibitem{Karapetyan:2018yhm} G. Karapetyan, Phys.\ Lett.\ B {\bf 786}, 418 (2018) [arXiv:1807.04540 [nucl-th]].

\bibitem{daRocha:2021imz}
R.~da Rocha,
%``Deploying heavier $\eta$ meson states: Configurational entropy hybridizing AdS/QCD,''
Phys. Lett. B \textbf{814}, 136112 (2021)
[arXiv:2101.03602 [hep-th]].

\bibitem{Fernandes-Silva:2018abr}
A.~Fernandes-Silva, A.~J.~Ferreira-Martins and R.~da Rocha,
%``The extended minimal geometric deformation of SU($N$) dark glueball condensates,''
Eur. Phys. J. C \textbf{78}, 631 (2018)
%doi:10.1140/epjc/s10052-018-6123-3
[arXiv:1803.03336 [hep-th]].

\bibitem{Ferreira:2020iry}
L.~F.~Ferreira and R.~da Rocha,
%``Nucleons and higher spin baryon resonances: An AdS/QCD configurational entropic incursion,''
Phys. Rev. D \textbf{101}, 106002 (2020)
[arXiv:2004.04551 [hep-th]].

\bibitem{Bernardini:2019stn}
A.~E.~Bernardini and R.~da Rocha,
%``Cosmological comoving behavior of the configurational entropy,''
Phys. Lett. B \textbf{796}, 107 (2019)
[arXiv:1908.04095 [gr-qc]].

\bibitem{Bazeia:2021stz}
D.~Bazeia and E.~I.~B.~Rodrigues,
%``Configurational entropy of skyrmions and half-skyrmions in planar magnetic elements,''
Phys. Lett. A \textbf{392}, 127170 (2021).

\bibitem{Braga:2023qee}
N.~R.~F.~Braga, L.~F.~Ferreira and O.~C.~Junqueira,
%``Configuration entropy of a rotating quark-gluon plasma from holography,''
Phys. Lett. B \textbf{847}, 138265 (2023)
%doi:10.1016/j.physletb.2023.138265
[arXiv:2301.01322 [hep-th]].


\bibitem{Braga:2021fey}
N.~R.~F.~Braga, Y.~F.~Ferreira and L.~F.~Ferreira,
%``Configuration entropy and stability of bottomonium radial excitations in a plasma with magnetic fields,''
[arXiv:2110.04560 [hep-th]].



\bibitem{Braga:2021zyi}
N.~R.~F.~Braga and O.~C.~Junqueira,
%``Configuration entropy in the soft wall AdS/QCD model and the Wien law,''
Phys. Lett. B \textbf{820}, 136485 (2021)
%doi:10.1016/j.physletb.2021.136485
[arXiv:2105.12347 [hep-th]].

\bibitem{Braga:2020hhs}
N.~R.~F.~Braga and R.~da Mata,
%``Configuration entropy description of charmonium dissociation under the influence of magnetic fields,''
Phys. Lett. B \textbf{811}, 135918 (2020)
%doi:10.1016/j.physletb.2020.135918
[arXiv:2008.10457 [hep-th]].


\bibitem{Braga:2016wzx} N. R. F. Braga and R. da  Rocha, Phys.\ Lett.\ B {\bf 767}, 386 (2017) [arXiv:1612.03289 [hep-th]].

\bibitem{Karapetyan:epjp} G. Karapetyan, Eur. Phys. J. Plus {\bf 136}, 1012 (2021) [arXiv:2105.07546 [hep-ph]].

\bibitem{Karapetyan:plb} G. Karapetyan, Phys.\ Lett.\ B {\bf 781}, 201 (2018) [arXiv:1802.09105 [hep-ph]].

\bibitem{Karapetyan:2020epl} G. Karapetyan, EPL {\bf 129}, 18002 (2020) [arXiv:1912.10071 [hep-ph]].

\bibitem{Karapetyan:2021epjp} G. Karapetyan, Eur. Phys. J. Plus {\bf 136}, 122 (2021) [arXiv:2003.08994 [hep-ph]].

\bibitem{Karapetyan:2023kzs}
G.~Karapetyan,
%``Configurational entropy and the N\ensuremath{*}(1440) Roper resonance in QCD,''
Annals Phys. \textbf{462}, 169612 (2024)
%doi:10.1016/j.aop.2024.169612
[arXiv:2305.05413 [hep-ph]].

\bibitem{Karapetyan:2021crv}
G.~Karapetyan,
%``Nuclear configurational entropy and high-energy hadron-hadron scattering reactions,''
Eur. Phys. J. Plus \textbf{137}, no.5, 590 (2022)
%doi:10.1140/epjp/s13360-022-02736-1
[arXiv:2112.11359 [nucl-th]].

\bibitem{Ferreira:2019inu} L. F. Ferreira and R. da Rocha, Phys.\ Rev.\ D {\bf 99}, 086001 (2019) [arXiv:1902.04534 [hep-th]].


\bibitem{Bernardini:2018uuy} A. E. Bernardini and R. da Rocha, Phys.\ Rev.\ D {\bf 98}, 126011 (2018) [arXiv:1809.10055 [hep-th]].


\bibitem{Braga:2018fyc} N. R. F. Braga, L. F. Ferreira and R. da  Rocha, Phys.\ Lett.\ B {\bf 787}, 16 (2018) [arXiv:1808.10499 [hep-ph]].


\bibitem{Bernardini:2016hvx} A. E. Bernardini and R. da  Rocha, Phys.\ Lett.\ B {\bf 762}, 107 (2016) [arXiv:1605.00294 [hep-th]].


\bibitem{Barbosa-Cendejas:2018mng} N. Barbosa-Cendejas, R. Cartas-Fuentevilla, A. Herrera-Aguilar, R. R. Mora-Luna and R. da  Rocha, Phys.\ Lett.\ B {\bf 782}, 607 (2018) [arXiv:1805.04485 [hep-th]].


\bibitem{Ferreira-Martins:2021cga}
A.~J.~Ferreira-Martins and R.~da Rocha,
%``Generalized Navier\textendash{}Stokes equations and soft hairy horizons in fluid/gravity correspondence,''
Nucl. Phys. B \textbf{973}, 115603 (2021)
[arXiv:2104.02833 [hep-th]].

\bibitem{Toniato:2025gts}
B.~Toniato, D.~Dudal, S.~Mahapatra, R.~da Rocha and S.~S.~Jena,
%``Holographic QCD model for heavy and exotic mesons at finite density: A self-consistent dynamical approach,''
[arXiv:2502.12694 [hep-th]].

\bibitem{Bernardini:2016qit} A. E. Bernardini, N. R. F. Braga and R. da  Rocha, Phys.\ Lett.\ B {\bf 765}, 81 (2017) [arXiv:1609.01258 [hep-th]].


\bibitem{Braga:2017fsb} N. R. F. Braga and R. da  Rocha, Phys.\ Lett.\ B {\bf 776}, 78 (2018) [arXiv:1710.07383 [hep-th]].


\bibitem{daSilva:2017jay} A. Goncalves da Silva and R. da Rocha, Phys.\ Lett.\ B {\bf 774}, 98 (2017) [arXiv:1706.01482 [hep-th]].

\bibitem{Guo:2024nrf}
X.~Guo, M.~A.~Martin Contreras, X.~Chen and D.~Xiang,
%``Holographic bottom-up approach to \ensuremath{\Sigma} baryons*,''
Chin. Phys. C \textbf{49}, 013104 (2025)
%doi:10.1088/1674-1137/ad7d75
[arXiv:2404.16608 [hep-ph]].

\bibitem{Colangelo:2018mrt} P. Colangelo and F. Loparco, Phys.\ Lett.\ B {\bf 788}, 500 (2019) [arXiv:1811.05272 [hep-ph]].


\bibitem{Karapetyan:2021ufz}
G.~Karapetyan and R.~da Rocha,
%``Configurational entropy of heavy-quark QCD exotica,''
Eur. Phys. J. Plus \textbf{136}, 993 (2021)
%doi:10.1140/epjp/s13360-021-01942-7
[arXiv:2103.10863 [hep-ph]].


\bibitem{MartinContreras:2023oqs}
M.~A.~Martin Contreras and A.~Vega,
%``Holographic stability for non-qq\textasciimacron{} candidates,''
Phys. Rev. D \textbf{108}, 126024 (2023)
%doi:10.1103/PhysRevD.108.126024
[arXiv:2309.02905 [hep-ph]].

\bibitem{MartinContreras:2022lxl}
M.~A.~Martin Contreras, A.~Vega and S.~Diles,
%``A holographic bottom-up description of light nuclide spectroscopy and stability,''
Phys. Lett. B \textbf{835}, 137551 (2022)
%doi:10.1016/j.physletb.2022.137551
[arXiv:2206.01834 [hep-ph]].


\bibitem{Ma:2018wtw} C. W. Ma and Y. G. Ma, Prog.\ Part.\ Nucl.\ Phys. {\bf 99}, 120 (2018) [arXiv:1801.02192 [nucl-th]].


\bibitem{daRocha:2021xwq}
R.~da Rocha,
%``Holographic entanglement entropy, deformed black branes and deconfinement in AdS/QCD,''
Phys. Rev. D \textbf{105},  026014 (2022) 
%doi:10.1103/PhysRevD.105.026014
[arXiv:2111.01244 [hep-th]].


\bibitem{daRocha:2020gee}
R.~da Rocha and A.~A.~Tomaz,
%``MGD-decoupled black holes, anisotropic fluids and holographic entanglement entropy,''
Eur. Phys. J. C \textbf{80} 857 (2020)
%doi:10.1140/epjc/s10052-020-8414-8
[arXiv:2005.02980 [hep-th]].

\bibitem{DaRocha:2019fjr}
R.~da Rocha and A.~A.~Tomaz,
%``Holographic entanglement entropy under the minimal geometric deformation and extensions,''
Eur. Phys. J. C \textbf{79}, no.12, 1035 (2019)
%doi:10.1140/epjc/s10052-019-7558-x
[arXiv:1905.01548 [hep-th]].


\bibitem{Gleiser:2012tu} M. Gleiser and N. Stamatopoulos, Phys.\ Rev.\ D {\bf 86}, 045004 (2012) [arXiv:1205.3061 [hep-th]].


\bibitem{Gleiser:2011di} M. Gleiser and N. Stamatopoulos, Phys.\ Lett.\ B {\bf 713}, 304 (2012).

\bibitem{Gleiser:2013mga} M. Gleiser and D. Sowinski, Phys.\ Lett.\ B {\bf 727}, 272 (2013) [arXiv:1307.0530 [hep-th]].


\bibitem{Gleiser:2018kbq} M. Gleiser, M. Stephens and D. Sowinski, Phys.\ Rev.\ D {\bf 97}, 096007 (2018) [arXiv:1803.08550 [hep-th]].

\bibitem{Gleiser:2018jpd}
M.~Gleiser and D.~Sowinski,
%``Configurational information approach to instantons and false vacuum decay in $D$-dimensional spacetime,''
Phys. Rev. D \textbf{98}, 056026 (2018)
%doi:10.1103/PhysRevD.98.056026
[arXiv:1807.07588 [hep-th]].

\bibitem{Sowinski:2015cfa} M. Gleiser and D. Sowinski, Phys.\ Lett.\ B {\bf 747}, 125 (2015) [arXiv:1501.06800 [hep-th]].


\bibitem{Gleiser:2014ipa} M. Gleiser and N. Graham, Phys.\ Rev.\ D {\bf 89}, 083502 (2014) [arXiv:1401.6225 [hep-th]].


\bibitem{Gleiser:2015rwa} M. Gleiser and N. Jiang, Phys.\ Rev.\ D {\bf 92}, 044046 (2015) [arXiv:1506.05722 hep-th]].

\bibitem{Meert:2020sqv}
P.~Meert and R.~da Rocha,
%``Probing the minimal geometric deformation with trace and Weyl anomalies,''
Nucl. Phys. B \textbf{967}, 115420 (2021)
%doi:10.1016/j.nuclphysb.2021.115420
[arXiv:2006.02564 [gr-qc]].

\bibitem{Abdalla:2009pg}
M.~C.~B.~Abdalla, J.~M.~Hoff da Silva and R.~da Rocha,
%``Notes on the Two-brane Model with Variable Tension,''
Phys. Rev. D \textbf{80}, 046003 (2009)
%doi:10.1103/PhysRevD.80.046003
[arXiv:0907.1321 [hep-th]].

\bibitem{Casadio:2016aum} R. Casadio and R. da Rocha, Phys.\ Lett.\ B {\bf 763}, 434 (2016) [arXiv:1610.01572 [hep-th]].


\bibitem{Fernandes-Silva:2019fez} A. Fernandes-Silva, A. J. Ferreira-Martins and R. da  Rocha, Phys.\ Lett.\ B {\bf 791}, 323 (2019) [arXiv:1901.07492 [hep-th]].

\bibitem{Braga:2020myi}
N.~R.~F.~Braga and R.~da Mata,
%``Configuration entropy for quarkonium in a finite density plasma,''
Phys. Rev. D \textbf{101}, no.10, 105016 (2020)
%doi:10.1103/PhysRevD.101.105016
[arXiv:2002.09413 [hep-th]].

\bibitem{Bazeia:2018uyg} D. Bazeia, D. C. Moreira and E. I. B. Rodrigues, J.\ Magn.\ Magn.\ Mater. {\bf 475}, 734 (2019) [arXiv:1812.04950 [cond-mat.mes-hall]].

\bibitem{Karapetyan:2019epl} G.  Karapetyan, EPL {\bf 125}, 58001 (2019) [arXiv:1901.05349 [hep-ph]].


\bibitem{Karapetyan:2017edu} G. Karapetyan, EPL {\bf 118}, 38001 (2017) [arXiv:1705.1061 [hep-ph]].


\bibitem{Karapetyan:2016fai} G. Karapetyan, EPL {\bf 117}, 18001 (2017) [arXiv:1612.09564  [hep-ph]].

\bibitem{Brower:2006ea}
R. C. Brower, J. Polchinski, M. J. Strassler, C.- I. Tan, JHEP {\bf 0712}, 005 (2007).

\bibitem{Watanabe}
A.~Watanabe and M.~Huang,
%``Total hadronic cross sections at high energies in holographic QCD,''
Phys. Lett. B \textbf{788}, 256 (2019)
%doi:10.1016/j.physletb.2018.11.042
[arXiv:1809.02515 [hep-ph]].



\bibitem{Correa:2015lla}
R.~A.~C.~Correa, R.~da Rocha and A.~de Souza Dutra,
%``Entropic information for travelling solitons in Lorentz and CPT breaking systems,''
Annals Phys. \textbf{359}, 198 (2015)
%doi:10.1016/j.aop.2015.04.027
[arXiv:1501.02000 [hep-th]].


\bibitem{Bonora:2014dfa}
L.~Bonora, K.~P.~S.~de Brito and R.~da Rocha,
%``Spinor Fields Classification in Arbitrary Dimensions and New Classes of Spinor Fields on 7-Manifolds,''
JHEP \textbf{02}, 069 (2015)
%doi:10.1007/JHEP02(2015)069
[arXiv:1411.1590 [hep-th]].

\bibitem{MarinhoRodrigues:2020yzh}
D.~Marinho Rodrigues and R.~da Rocha,
%``Odd-spin glueballs, AdS/QCD and information entropy,''
Phys. Lett. B \textbf{811}, 135943 (2020)
%doi:10.1016/j.physletb.2020.135943
[arXiv:2009.01890 [hep-th]].

\bibitem{MarinhoRodrigues:2020ssq}
D.~Marinho Rodrigues and R.~da Rocha,
%``Configurational entropy and spectroscopy of pomeron resonances in dynamical AdS/QCD,''
Eur. Phys. J. Plus \textbf{137}, 429 (2022)
%doi:10.1140/epjp/s13360-022-02622-w
[arXiv:2006.00332 [hep-th]].


\end{thebibliography}
\end{document}